\documentstyle[12pt,aasms4]{article}

\input epsf.sty
\begin{document}

\title{The Blazar Main Sequence}

\author{A. Cavaliere, V. D'Elia}

\affil{Astrofisica, Dip. Fisica 2a Univ. Tor Vergata,  Roma
I-00133}

\baselineskip=22pt
\def\lsim{\ \raise -2.truept\hbox{\rlap{\hbox{$\sim$}}\raise5.truept
\hbox{$<$}\ }}
\def\gsim{\ \raise -2.truept\hbox{\rlap{\hbox{$\sim$}}\raise5.truept
\hbox{$>$}\ }}
\def\msun{M_{\odot}}
\def\mincir{\ \raise -2.truept\hbox{\rlap{\hbox{$\sim$}}\raise5.truept
\hbox{$<$}\ }}
\def\magcir{\ \raise -2.truept\hbox{\rlap{\hbox{$\sim$}}\raise5.truept
\hbox{$>$}\ }}

\begin{abstract}

We propose a sequence (the Blazar main sequence, BMS) that links
the two main components of the Blazar class, namely, the Flat
Spectrum Radio Quasars and the BL Lacertae objects, and yields all
their distinctive features in a correlated way. In this view, both
type of sources are centered on a supermassive Kerr hole close to
maximal spin and observed pole on. But the FSRQs are energized by
accretion at rates $\dot m \sim 1 - 10$, and are dominated by the
disk components (thermal and electrodynamic jet-like component)
which provide outputs in excess of $L \sim 10^{46}$ erg s$^{-1}$.
On the other hand, accretion levels $\dot m \ll 1$ are enough to
energize the BL Lacs; here the radiation is highly non-thermal and
the power is partly contributed by the rotational energy of the
central Kerr hole, with the latter and the disk together
sustaining typical $L \sim 10^{44}$ erg s$^{-1}$ for several Gyrs.
If so, we expect the BL Lacs to show quite different evolutionary
signatures from the FSRQs, and in particular, number counts close
to the Euclidean shape, or flatter if the sources make a
transition to the BL Lac from a FSRQ mode. In addition, for lower
$\dot m$  along the BMS we expect the large scale electric fields
to be less screened out, and to accelerate to higher energies
fewer particles radiating at higher frequencies; so in moving from
FSRQs to BL Lacs these non-thermal radiations will peak at
frequencies inversely correlated with the disk output. For the BL
Lacs such dependence implies increased  scatter when one tries a
correlation with the total ouput. At its endpoint, the BMS
suggests widespread objects that are radiatively silent, but still
efficient in accelerating cosmic rays to ultra high energies.

\end{abstract}


\section{Introduction}

The Blazars are commonly perceived as a class of Active Galactic
Nuclei marked out from the rest of their kin by a number of
features: strong, high frequency radio emission from compact
cores, often with superluminal expanding components (see Jorstad
et al. 2001); powerful $\gamma$ rays extending to energies $h \nu
\sim 10^2$ MeV and beyond (see Mukherjee 1999); rapid
multi-wavelength variability (see B\"ottcher 2000); high and
variable optical polarization (see Yuan et al. 2001).
Comprehensive reviews and extensive references concerning these
sources may be found in Urry \& Padovani (1995), Padovani \& Urry
(2001, henceforth PU01).

The above features defining the Blazar {\it class} are widely
explained in terms of the view originally proposed by Begelman,
Blandford \& Rees (1984). This holds the radiation from all these
sources to be produced in a collimated, relativistic jet of
particles with bulk Lorentz factor $\Gamma \sim 5 - 20$; when
observed at small angles of order $\Gamma^{-1}$ the jet produces
the ``blazing'' effects that provide the class denomination. The
intrinsic luminosity emitted by the jet over the entire solid
angle, and total jet kinetic power are reviewed by Ghisellini
(1999a), whose definitions and beaming factors we will adopt in
the following.

Within the Blazar class two main {\it subclasses} are usually
identified, namely,  the BL Lac Objects and the Flat Spectrum
Radioloud Quasars (FSRQs). The former differs from the latter
subclass on several accounts, that we shall discuss in this paper:
a) lack of emission lines with EW $\magcir 5$ A, and lack of
blue-UV bumps; b) total outputs  $L \mincir 10^{46}$ erg s$^{-1}$,
compared with FSRQs often exceeding this limit and in some cases
approaching $10^{48}$ erg s$^{-1}$ (see specifically Maraschi
2001); c) spectral energy distributions peaking at frequencies
from optical to X-ray and in the range of $10$ GeV, compared with
the FSRQ peaks at far IR to optical frequencies and around
$10^{-1}$ GeV; d) no signs of cosmological evolution, compared
with the strong evolution clearly displayed by the FSRQs in common
with the other quasars.

These distinctive features stand out of the many selections
entangling this field, and even though {\it intermediate} objects
are currently found (Sambruna et al. 2000, Perlman et al. 2001),
the FSRQ/BL Lac strong bimodality calls for explanation; this is
the scope of the present work, that expands some preliminary
results reported by D'Elia \& Cavaliere (2001).

We will base on the  accreting black hole (BH) paradigm (see
Begelman, Blandford \& Rees 1984), in the framework of a
4-parameter classification similar to that discussed by Blandford
(1990, 1993); this comprises the BH mass $M$, its angular momentum
$J$, the accretion rate $\dot M$, and the viewing angle $\theta$
from the axis of the jet (if any). In the scheme we contemplate,
radio quiet AGNs correspond to slowly rotating BHs with low
(Seyfert galaxies) or full (quasars) $\dot M$ compared to the
Eddington rates; the quasars require large BH masses $M$ of order
$10^9\, M_{\odot}$ to attain proportionally high Eddington
luminosities $L_E \approx 10^{47}\,M_9$ erg s$^{-1}$.

Radio loudness constitutes a debated issue. Radio loud AGNs are
often observed to reside in very bright galaxies. On the one hand,
these sources have been inferred to correlate with large BH
masses; for instance, Laor (2000) stresses the case for radio
loudness being a consequence of large $M$. On the other hand, the
majority of the optically selected quasars including very bright
ones are radio quiet (see for recent data Goldschmidt at al. 1999,
Hamilton et al. 2001); in addition, Scarpa \& Urry 2001 find that
the hosts of quasars strong in the radio are no different from
giant ellipticals in general, including hosts of radio quiet
quasars. Finally, Ho (2001) from a comprehensive database finds no
simple relation between radio loudness and $M$. Similar data from
more limited observations spurred the view (see Blandford 1993)
that the other BH parameter, namely the spin, should be involved
as a necessary, intrinsic condition for directing and launching
the jets associated with radio emissions. In the present paper we
focus on this view.

In fact, we will consider the Blazars in the context of
supermassive holes with  $M  \sim 10 ^9 M_{\odot}$ close to the
maximal observed values for MDOs (see Gebhardt et al. 2000,
Ferrarese \& Merritt  2000), with high  angular momentum $J \sim
J_{max}$ close to the maximal $J_{max} = GM^2/c$, observed nearly
pole on within angles $\theta \mincir \Gamma^{-1}$ (see Sambruna,
Maraschi \& Urry 1996). We shall concentrate on discussing how
{\it all} Blazar properties can be arranged into a Blazar main
sequence (BMS) in terms of levels of the accretion rate $\dot m =
\dot M\,c^2/L_E$. Levels $\dot m \sim \eta^{-1} \sim 1 - 10$ (the
radiative efficiency being up to about 0.4 around a Kerr hole)
mark the FSRQs, while values $\dot m \mincir 10^{-2}$ mark the BL
Lacs.

Our plan is as follows. In \S 2 we will stress how the optical
properties of the Blazars can be understood in terms of different
levels of $\dot m$. In \S 3 we show that high rates $\dot m \sim
1$ can produce the huge thermal and non-thermal outputs featured
by many FSRQs, while conditions where $\dot m \ll 1$ lead to the
moderate, but mainly non-thermal outputs of the BL Lacs. In \S 4
we will argue that for lower values of $\dot m$ along the BMS the
typical particle and photon energies are expected to increase; a
secondary physical parameter relevant at low  $\dot m$ causes
considerable variance within the BL Lac subclass, adding to the
effects from the  orientation. In \S 5  we show how in the same
framework we understand the widely different evolutionary
properties observed in FSRQs and in BL Lacs. Finally, in \S 6 we
summarize and discuss our conclusions.

\section{Optical Properties}

In the standard view (see Frank, King \& Raine 1992; Peterson
1997) all {\it thermal} outputs are  produced in, or around  the
accretion disk, with only weak anisotropies and with power given
by $ L_{th} \approx \eta \,\dot m \,L_{E}$.

Here we take up this view, and just stress how from the broad
optical emission lines and the blue-UV bump in particular we
derive a first indication toward a BMS in terms of levels of $\dot
m$. The bump is produced by power reprocessed in a superposition
of black body emissions from the inner rings of the accretion
disk, or in a hot corona (see Sun \& Malkan 1989, Siemoginowska et
al. 1995); the lines are produced
 by the high UV-soft X-ray continuum
reprocessed by distant gas ``clouds'' (see Netzer 1990). In both
cases the density of the reprocessing material is expected to
scale up with increasing $\dot m$.

The FSRQs share with other quasars conspicuous bumps and prominent
broad emission lines; the strength of these features are
consistent with high accretion rates $\dot m \sim 1$.

On the other hand, the weakness or absence of such features in the
BL Lac spectra can be understood in terms of $\dot m \mincir
10^{-2}$. In fact,  at such  levels of $\dot m$ not only a weaker
ionizing/exciting continuuum is emitted by the disk, but also
lower gas densities are expected in the disk and its environment
out to pc and larger scales; these conditions concur to account
for the weak or absent emission lines.

The other effect that in the BL Lacs acts to swamp any residual
bump and to further reduce the EW of intrinsically weak emission
lines
 is constituted by the blazing continuum from the jet when observed at a small
angle, see Vagnetti, Giallongo \& Cavaliere (1991); Georganopoulos
(2000).

We recall that low values of $\dot m$ related to radiosources had
been proposed by Begelman et al. (1984). Specifically, the
different power levels and optical spectra of the FRII and FRI
radiosources have been previously related to high and low values
of $\dot m$, respectively, by Baum, Zirbel, {\&} O'Dea 1995; this
point will be further discussed in \S 6.

\section{The Blazar Power}

As to the {\it jets}, we assume them to be powered by variants of
the mechanism  originally proposed by Blandford \& Znajek (1977,
BZ) for direct extraction of rotational energy from a Kerr hole
via the Poynting-like flux associated with the surrounding
magnetosphere. The BZ power scales as
$$ L_K \propto B_h^2 \; r^2_h ~, \eqno (1)$$
in terms of the magnetic field $B_h$ threading the hole horizon at
$r_h$.

Variants are necessary in view of the limitations to $L_K$
recently stressed  by Modersky \& Sikora (1996); Ghosh \&
Abramowicz (1997); Livio, Ogilvie \& Pringle (1999). Such variants
involve either high strengths of $B_h$, or the similar
electrodynamic output of the disk from a larger radius $r_{d}$.
%larger than $r_h$.

In the first variant, very strong $B_h$ as advocated by Meier
(1999) are required to account for the huge outputs of some FSRQs.
Fields up to $B_h^2/8\pi\sim \rho c^2$ in the plunging orbit
region have been argued by Krolik (1999); Armitage, Reynolds \&
Chiang (2000) and Paczynski (2000) discuss why such field values
are unlikely in a thin disk. In thick disks the status of such
enhanced fields is still unsettled.

The second variant takes up from recent discussions  (Ghosh \&
Abramowicz 1997; Livio, Ogilvie \& Pringle 1999) that have
stressed the continuity of $B_h$ with the field $B_d$ at the inner
rim of a standard $\alpha$-disk; in turn, $B_d$ is bounded by the
maximum pressure, following $B^2_d/4\pi <  P_{max}$. Depending on
$\dot m$, the inner disk is dominated by gas pressure (GPD) or by
radiation pressure (RPD); the pressure scales as $P_{max} \propto
(\alpha M_9)^{-9/10}{\dot m}_{-4}^{4/5}$ in the first,  and as as
$P_{max} \propto (\alpha M_9)^{-1}$ in the second regime.
Correspondingly, in the first regime the  maximum power
extractable from the {\it hole} grows with $\dot m$, and in the
jet frame the emission reads (see Modersky \& Sikora 1996; Ghosh
\& Abramowicz 1997)
$$L_{K}= 10^{44}M_9^{11/10}\dot m_{-4}^{4/5}(J/J_{max})^2 \; erg \,s^{-1}~~~~~~
(GPD)~ \eqno (2)$$ in terms of the Kerr hole mass $M= M_9 \, 10^
9\; M_{\odot}$; but for $\dot m \magcir 10^{-3}$ the power $L_K$
saturates to
$$ L_{K}=2 \, 10^{45}\,M_9 \,(J/J_{max})^2 \;  erg \, s^{-1}~~~~~~ (RPD)~.
\eqno(3) $$
This is because the radius $r_c$ defining the RPD region grows
with $\dot m^{16/21}$ (Novikov \& Thorne 1973), and it turns out
to exceed the radius $r_{ms}$ of the last stable orbit when $\dot
m \magcir 10^{-3}$ holds; in the latter conditions an inner RPD
region actually exists.

On the other hand, the electrodynamic power $L_d$ contributed by
the {\it disk} and itself strongly anysotropic scales as
$$ L_d \propto B_d^2 \; r^2_d ~; \eqno (4)~$$
here the effective disk radius  $r_d$ is bounded at several times
$r_h$ either by the value of $r_c$ or by the radial decline in the
disk of the efficiency $\eta \sim G\,M\,\dot M/2\, r$. We recall
that the ratio  $L_d / L_K$ had originally been estimated by
Blandford \& Znajek 1977 at values  $\magcir 5$; recent
reappraisals (e.g., by Livio, Ogilvie \& Pringle 1999) have
evaluated it at $L_d/L_K \sim (B_d/B_h)^2  \;  (r_{d}/r_h)^{3/2}$.

We draw the following implications concerning the BMS. Many BL Lac
jets can live on accretion rates $\dot m \sim  10^{-2}$, since the
power levels $L_K$ given by eq. (3) are adequate even including
the kinetic energy in the jets (Celotti, Padovani \& Ghisellini
1997, Maraschi 2001). A larger but comparable contribution comes
from the disk, so that $ L_K \sim L_d  \magcir L_{th}$ holds. The
disk is coupled to the rotating hole both dynamically and
magnetically (see Livio, Ogilvie \& Pringle 1999) through the
magnetic field lines threading the hole and tethered in the disk.

On the other hand, many FSRQs feature total outputs $L > 10^{46}$
erg s$^{-1}$, with specific sources exceeding $10^{47}$erg
s$^{-1}$, see Tavecchio et al. (2000), and Maraschi (2001), who
include a high proton contribution to the jet energy and adopt
average values of $\Gamma \approx 10$ (see Sikora 2001). Such
luminosities require not only a saturated $L_K$ (see eq. 3), but
also a dominant component up to $L_d \sim 10^2\, L_K$ from the
disk corresponding to $B_d$ a few times larger than $B_h$  and
$r_d \sim 5 r_h$; this implies an extended region dominated by
radiation pressure, hence  conditions where $\dot m \sim 1$,
consistent with the requirements from the optical properties
discussed in \S 2.

We note that in FSRQ conditions both $L_d$ and the thermal
emission $L_{th}$ are fed by the gravitational power supply of
order $G\,M\,\dot M/2\, r_{d}$. The share between $L_d$ and
$L_{th}$ depends on the power spectrum of the magnetic field
inhomogeneities, that is, on the power share between two scales:
the large scale, coherent  vs. the small scale, turbulent
component. A dominant share of the former, such as to yield a
large $L_d$ and a large associated transfer of angular momentum
outward, may require modifications of the standard $\alpha$-disk
model as discussed by Salvati (1997).

The hole output $L_K$, even when it constitutes a minor component
to the total, is likely to be important to provide on the jet axis
a ``high-velocity spine'' instrumental for the outward jet
propagation, see Livio (2000), Chiaberge et al. (2000).

\section{The Blazar SEDs}

Another feature of the Blazars is constituted by their extended
spectral energy distribution (SED). This shows a first peak
commonly interpreted as synchrotron radiation (see Urry \&
Padovani 1995). In the FSRQ spectra the peak frequencies
$\nu_{peak}$ lie mainly in the range from the far IR to the
optical band; in the BL Lacs the peaks are
 generally  shifted to higher
frequencies, and in some objects up to the hard X rays (Costamante
et al. 2001).

An additional high energy component to the radiation (see PU01) is
observed into the 10 GeV range for the FSRQs, and out to several
TeVs for the BL Lacs \footnote{Note that the $\gamma$-ray spectra
at the sources may extend to higher energies than observed at
Earth after the intervening absorption due to the IR background.
If so, the following argument concerning coherent acceleration
would be reinforced}. This is likely to be inverse Compton
emission produced by substantial numbers of GeV or of $10^2-10^3$
GeV electrons, respectively, scattering off external or
synchrotron seed photons as discussed by Ghisellini (1999a).

The apparent inverse scaling of $\nu_{peak}$ with the radiative
$L$ in moving from FSRQs to (quiescent) BL Lacs has been pointed
out by Fossati et al. (1998), and interpreted as a cooling
sequence due to the radiative energy losses being faster in the
more powerful sources, up to controlling the maximal electron
energies. Concerning the acceleration, one view considers weak
electric fields over relatively large distances; these occur in
the internal shock scenario where, however, the typical energies
produced are of order ${\cal E}\sim 2\, \Gamma \, m_p \, c^2 < 20$
GeV and fall short of the top energies required in some BL Lacs as
discussed by Ghisellini (1999b).

So we are lead to consider as an addition or as an alternative
particularly relevant to BL Lacs, the electron acceleration due to
higher electric fields acting over shorter distances. We shall see
that this leads to an {\it acceleration} sequence also linked with
the BMS; this is because in the BL Lacs lower levels of $\dot m$
allow for higher electron energies and so produce a SEDs peaked at
higher frequencies. Meanwhile, a 2$^{nd}$ intrinsic parameter
produces the {\it specific} scatter of $\nu_{peak}$ within this
subclass.

Such high $E$ fields naturally arise around the BZ magnetosphere
in association  with considerable energy transport via ``Pointing
flux'' along the jet, with the magnetic field decreasing outwards
rather slowly, like $B =  B_d \, (r_{ms}/r)^{(1+p)}$ with $p < 1$;
typically $p = 1/4$, in a similar manner to Blandford {\&} Payne
(1982). The force-free condition $ E \bullet B = 0$ governing the
BZ magnetosphere has to break down at the flow boundaries, and to
give way to electric fields $E \mincir  B$; this is likely to
occur time-dependently and inhomogeneously within sheets or
filaments located at average distances $r \sim 10^{16} \div
10^{18}$ cm. But such $E $ fields are electrodynamically screened
out beyond distances exceeding a few times $ d = c/\omega_p
\propto (\gamma / n)^{1/2}$, where  $\omega_p = (4 \pi e^2
n/m\gamma)^{1/2}$ is the plasma frequency of the screening
particles with Lorentz factor
 $\gamma$ and density  $n$.

Lower bounds to the densities may be estimated from the powers
emitted in the form of synchrotron or inverse Compton radiation,
that is, $L \approx \gamma^2 \,U\, R^3\,n$ where $U$ is the total
energy density in quiescent conditions. We obtain $n \sim 1$
cm$^{-3}$ for BL Lacs with $L \sim 10^{44}$ erg s$^{-1}$, and $n
\sim 10^{3}$ cm$^{-3}$ for FSRQs with $L \sim 10^{47}$ erg
s$^{-1}$. Correspondingly, we expect many electrons to attain high
energies, up to
$$\gamma_{max} \mincir e\, B\, d \,  /m_e\, c^2 \approx 10^8 \, B_4 \,
M^{1.25}_9 \, d_{10} \, r^{-1.25}_{17}  ~, \eqno (5) $$
where we have used $d = 10^{10} \, d_{10}$ cm, and $r = 10^{17}\,
r_{17}$ cm; we have also expressed as $B_d = 10^4 \; B_4$ G  the
field values at $r_{ms} \approx 1.5\, 10^{14}\, M_9$ cm, which
decrease into the emitting region like $B(r) = B_d \,
(r_{ms}/r)^{1.25}$ as recalled above.

In eq. (5) there is room for variations of $B$ and $n$; the former
may be down by a factor 10 or  the latter up by factors $10^2$
(with $M$ again set at $10^9 \, M_{\odot}$), and eq. (5) still
yields $\gamma \sim 10^7$. This is enough to produce inverse
Compton photons up to $\sim 10$ TeV, the highest energy photons
observed to now from the BL Lacs. In FSRQs densities $10^3$ larger
give electron energies some 30 times smaller, consistently with
the parameters derived by B\"ottcher \& Dermer 2001 from advanced
spectral modeling. Given $\gamma_{max}$, we expect less energetic
but more numerous electrons to arise naturally when the particles
escape before full acceleration or $E < B$ holds, and when the
inhomogeneities of $B$ are considered; the latter varies outwards
following $B \propto r^{-1.25}$ as said, and this condition by
itself produces  an energy range by factors close to $10^{3}$
within a distance range $r \sim 10^{17 \pm 1}$ cm.

The electron energies given by eq. (5) are proportional to  $d
\propto n^{-1/2}$, and so scale like
$$\gamma_{max} \propto
(\gamma^2 UR^3)^{1/2} \;L^{-1/2}. \eqno (6)$$
This results in $\gamma_{max} \propto L^{-1/2}$ if we adopt the
empirical relation $\gamma^2 U \sim$ const reported by Ghisellini
(1999a) for the FSRQs. On the other hand, the same author finds
that the $\gamma $ values for the BL Lacs exceed this relation,
but may be included on modifying it to read $\gamma^2 U \sim
U^{-0.2}$; from this we obtain $\gamma_{max} \propto L^{-3/5}$.
Thus we expect the peaks of the non-thermal emissions, and more
clearly the syncrotron $\nu_{peak} \propto  \gamma^2_{max}$, to
anticorrelate strongly with the disk luminosity following
$\nu_{peak} \propto L^{-6/5}$ or steeper, when comparing the FSRQ
with the BL Lac subclass.

In closer detail, the emitting particles are likely to constitute
the high energy subset of the total particle population; but
similar results obtain if all electrons are described as in
Ghisellini (1999a) by a power-law energy distribution $n(\gamma)
\propto \gamma^{-a}$ from $\gamma_{min} \sim 10^3$; with $a
\approx 2.5$ we obtain $\nu_{peak} \propto \gamma^2_{max} \propto
L^{-4/3}$, while  a steeper relation $\nu_{peak} \propto  L^{-2}$
holds with $a \approx  2$. Concerning the full SED we refer to the
complete modeling of the emitting regions by B\"ottcher \& Dermer
2001, who treat quantitatively the spectral shapes radiated by
synchrotron and inverse Compton emissions in the jets, considering
also the cooling limit to the electron energies provided by the
high photon densities in FSRQs; they evaluate specific parameters
agreeing with the orders of magnitude we estimate.

The other outcome from our view is that within the BL Lac subclass
we also expect the {\it second} parameter $L_K/L_d$ to affect the
scaling laws provided by eq. (6). This is because the total
radiative luminosity of the BL Lacs $L = L_d + L_K$ comprises a
disk and a Kerr hole component with comparable powers. Of these,
only $L_d \sim G\, M\, \dot M / 2 \, d_{ms}$ is directly related
to  $\dot m$ and to $n$ while $L_K$, when considerable, tends to
be independent of them as shown by eq. (3). Thus at a given $L_d$
(hence at given $\dot M$ and $n$) higher total luminosities obtain
in BL Lacs where $L_K$ is important. In other words, $\nu_{peak}$
scales  inversely with the component $L_d$ only, and it is bound
to show scatter {\it specific} to the BL Lacs when one tries to
correlate it  with the total luminosity. As a consequence, we do
not expect the scaling to extend unblurred from the so called LBLs
to the HBLs (low and high energy peaked BL Lacs, respectively, see
Padovani \& Giommi 1995). In fact, considerable scatter of
$\nu_{peak}$ at given $L$ is increasingly found within the BL Lac
subclass, see Giommi et al. (2001).

On the other hand, for the FSRQs $L \approx L_d$ holds, that is,
$L_K $ is irrelevant as to the total power, and so is its
contribution to the scatter.

The specific BL Lac scatter discussed here adds to other
components common to all Blazars due to different values of $B,
M$, and viewing angles (see Sambruna, Maraschi \& Urry 1996,
Georganopoulos 2000), and to variations of $\nu_{peak}$ and $L$
arising during flares.

To this point we have shown how all spectral properties of FSRQs
and BL Lacs are expected to consistently vary along the BMS. In
the next section we discuss why we expect the Blazar evolutions to
be also related to different levels of $\dot m$.

\section{The Blazar Evolutions}

Strong cosmological evolution is closely shared by the FSRQs with
the rest of the quasars (see Wall \& Jackson 1997, Goldschmidt et
al. 1999), and shows up, e.g., in their steep number counts; the
BL Lacs, instead, show no signs of a similar behavior (Giommi,
Menna \& Padovani 1999; Padovani 2001).

Here we argue that such an evolutionary peculiarity is related to
the following two independent circumstances: i) At levels $\dot m
\mincir 10^{-2}$ all BL Lacs shine for several Gyrs with only a
mild decrease in luminosity (a very slow ``luminosity evolution'',
LE), and this implies counts close to Euclidean. ii) A genetic
link is conceivable from FSRQs to BL Lacs, that is, some of the
former may switch from a short-lived regime of high $\dot m$ to
the long-lived regime of low $\dot m$; so the BL Lacs undergo also
some negative ``density evolution''(negative DE), and their counts
are further flattened. Since point i) has been discussed by
Cavaliere \& Malquori 1999, we just recall their results below,
and concentrate on point ii).

The evolution of the bright optically selected (and mainly radio
quiet) quasars is strong, see Boyle et al. (2000), and so is in
X-rays, see Della Ceca et al. (1994); Miyaji, Hasinger \& Schmidt
(2000). For the bright sources, this behavior is widely traced
back to the exhaustion in the host galaxy of the circumnuclear gas
stockpile usable for further accretion onto the central black
hole. Such an exhaustion is caused by previous accretion episodes
and by ongoing star formation (Cattaneo, Haehnelt \& Rees 1999;
Kauffmann \& Haehnelt 2000; Cavaliere {\&} Vittorini 2000, CV00);
the ensuing decrease of the average accretion rate $\dot m$ is
fast, and causes the average bolometric luminosities to scale down
rapidly for $z < 2.5$ (a strong LE). The radio loud quasars share
this evolutionary behavior, constituting a fraction (not yet
quantitatively understood, see \S 1, and the discussions by
Moderski, Sikora \& Lasota 1998 and by Livio 2000) of order 10\%
of the total; the FSRQs in particular are widely held to be the
same sources seen pole on.

But as the rates $\dot m$ decrease we also expect a considerable
fraction of the underlying hole-disk systems to eventually switch
from being overwhelmingly fueled by accretion to being partly fed
by the hole rotational supply $E_K$; the latter had been
stockpiled during the phases of accretion of coherent angular
momentum $J$ along with the mass (Bardeen 1970; Moderski, Sikora
\& Lasota 1998). In other words, we expect some radio loud sources
to switch from the FSRQ to the BL Lac mode. The moderate outputs
typical of this mode can be sustained for several Gyrs by the
dynamically and magnetically coupled system constituted by Kerr
hole and inner disk; so the decrease of BL Lac luminosities is
expected to be slow (Cavaliere {\&} Malquori 1999), with time
scales in the range estimated to a first approximation from
$\tau_L \sim E_{K}/L_K \approx 5 \div  10$ Gyr, rather uniform
within the population.

In closer detail, CV00 (see also refs. therein) trace  the
accretion episodes feeding the quasars back to interactions of the
host galaxies with their companions  in a group. These events
destabilize the host gas, and trigger accretion over times of
order $\tau \sim$ a few $10^{-1}$ Gyr that comprise the duration
of a close interaction and a few galactic dynamical times for the
inflow to develop and subside; some $3 - 5$ repetitions are
expected per host galaxy since $z \approx 2.5$. The frequency of
such interaction episodes diminishes in time, and gives rise to a
positive, weak DE of the quasar population over scales $\tau_D
\sim $ several Gyrs. Meanwhile, the efficiency of such episodes
drops as anticipated above, due to the exhaustion of the host gas;
the latter is halved on the scale $\tau_L \approx 3$ Gyr. The
result is that the average luminosities drop over the same scale,
in other words, a strong LE is produced in the quasar population.

On this basis CV00 compute the luminosity function (LF) for the
optically selected objects at $z<2.5$ to read
$$N_F (L,z) \approx
{\tau \,N_G(z) \over \tau_r(z) \, L_b(z)}\;\; {1 \over l^{1.2} +
l^{3.2}} ~. \eqno (7) $$
Here $N_G(z)$ is the space density of groups of galaxies,
$\tau_r^{-1}(z) \propto (1+z)^{3/2}$ is the average interaction
frequency for a host galaxy in a group, and $l\equiv L/L_b(z)$ is
normalized using the break luminosity that scales as
$L_b(z)\propto (1+z)^3$ in the critical universe; we use this
simple scaling for numerical evaluations, since CV00 show that the
observables do not vary much in the $ \Omega_o + \Omega_{\lambda}
= 1$ cosmology. For the FSRQs we will use here eq. (7) with the
proportionality factor of order $5\; 10^{-2}$.

Our main point here is that the episodes of powerful FSRQ activity
feeding on high $\dot m$ will last some $10^{-1}$ Gyr and die out
over similar scales; the sources may be no longer reactivated
after a ``last interaction''. This leaves behind a rapidly
spinning hole (and a relatively quiescent though bright galaxy,
see Scarpa 2001) in about 1/2 of the objects. Thus the DE
timescale $\tau_D$ for the deaths of the bright FSRQs also
constitutes the timescale for BL Lac births; we evaluate it from
eq. (7) integrated over $L\geq L_b$, to obtain
$$\tau_D^{-1}={d \over dt}\, ln \,  [ \,\int_1^{\infty} N(l,t)\, dl \, ] =
{d \over dt}\, ln[\,  N_G(t)/\tau_r(t)\, ] ~, \eqno (8)$$ with the
second equality holding because only the time-dependent amplitude
in eq. (7) is relevant here. We obtain values of $\tau_D$ between
7 and 5 Gyr, depending on the form for small $z$ of $N_G(z)$ which
may scale as $(1+z)$ on assuming a Press \& Schechter (1974) mass
distribution, or as $(1+z)^2$ following Lacey \& Cole (1993).

One sign of evolution is provided by the integrated source counts,
which may be evaluated from the relation (see Cavaliere \&
Maccacaro 1990)
$$ N(>S) \propto
S^{-3/2}[1-C(S_0/S)^{1/2} + 0(S^{-1})] ~.\eqno (9)$$
This simple relation is limited to high-intermediate fluxes. On
the other hand, it has the advantage of exposing how the
evolutionary properties of the two Blazar subclasses depend on the
two population time scales $\tau_D$ and $\tau_L$ (the latter being
long for the BL Lacs, and short for the FSRQs); these appear in
the coefficient
$$ C = 3\,D_0\; \langle l^2 \rangle [2(1+\alpha) -  1 /H_0\, \tau_D
-  (\beta-1) / H_0\, \tau_L]/4\,R_H\langle l^{3/2} \rangle~.
\eqno(10)$$

For the BL Lacs counted in the radio band a relevant value is
 $\tau_L\sim 10$ Gyr (marking the slow LE of the BL Lacs). Other quantities
involved are: the flat spectral index $\alpha= - 0.3$ in the GHz
range; the slope $\beta \approx 2.5$ of the radio LF; the
normalized moments $\langle l^n\rangle$ of the LF; the typical
distance to high flux BL Lacs $D_0 = (L_0/4 \pi S_0)^{1/2}
\approx 0.03 R_H$ in Hubble units.

On using only the slow LE on the scale $\tau_L \approx  8$ Gyr,
eq. (10) with the Hubble constant $H_0 = 65$ km/s Mpc
 yields $C= - \, 0.1$; then
from eq. (9) the result is $N(> F) \propto F ^{-1.6}$ as in
Cavaliere \& Malquori (1999).

But BL Lac births imply also {\it negative} DE to occur, described
by the scale $\tau_D \sim -\, 6$ Gyr evaluated below eq. (8). When
also this scale is introduced in eq. (10), the values of $C$
increases to $C\approx 0.1$. Now the result is $N(>S) \propto
S^{-1.5}$ at high-intermediate fluxes, consistent with the data by
Giommi et al. (1999), see fig. 1; at 5 GHz similarly flat counts
are both expected (see Jackson \& Wall 2001) and observed (see the
preliminary data by Padovani 2001), while the FSRQ counts observed
are quite steep (see Wall \& Jackson 1997,  Jackson \& Wall 2001).
We add that no evidence of evolution has been observed by
Caccianiga et al. (2001) from the values of $V_e/V_a$  in a large
sample of BL Lacs from the REX survey.

The count slopes we evaluate for the BL Lacs are in sharp contrast
with the result from a similar evaluation for the FSRQs. The
parameters appropriate for them are recalled above and read:
$\tau_D = +\, 6$ Gyr, $\tau_L = 3$ Gyr, while $ D_0 \approx 0.3\,
R_H$ holds and again $\beta \approx  2.5$. Eq. (10) then gives  $C
\approx - 8 $, and in turn eq. (9) yields radio counts $N(>S)$
steep well above $ S^{-1.5}$ and consistent with the data, see
fig. 2.

We thus see that also the evolutionary properties vary
systematically along the BMS. Actually, the faint end of the LF is
expected to be flattened by the varying beaming effects associated
with a distribution of $\Gamma$ values (Urry {\&} Padovani 1995);
this holds both for the FSRQs and the BL Lacs, and will contribute
to the flattening of the counts at their respective faint ends,
but it does not affect the intermediate range where sharply
different slopes stand out.

\section{Discussion and Conclusions}

We have argued that Blazars constitute  basically {\it similar}
sources comprising a supermassive, nearly maximally spinning Kerr
hole and its accretion disk; but large, systematic and correlated
changes are bound to arise in all observational properties when
the accretion rate levels vary from $\dot m \sim  1 $ that mark
the FSRQs to $\dot m \ll 1$ that mark the BL Lacs. Such systematic
changes basically arise  because lower values of $\dot m$ imply
{\it depleted} densities in and around the disk; by the same
token, these cause both {\it higher} energies of the emitting
particles with {\it less} reprocessing of the emitted photons, and
{\it lower} total luminosities over longer time scales leading to
{\it less} evolution.

Our view (in short, BMS for Blazar main sequence) outlined here
lead us to specifically predict the following {\it correlated}
features, also summarized in Table 1.

a) Optical properties: in \S 2 we recall that high values of $\dot
m$ yield strong thermal features, namely, emission lines and big
blue bump.  On the other hand, we stress that low values of $\dot
m$ imply intrinsic weakness or drowning of all these features,
especially in the presence of a strong beamed continuum.

b) Luminosities: in \S 3 we argue -- in view of the limitations to
the power directly extractable from a rotating hole -- that disks
with high values of $\dot m$ are required to produce outputs as
strong as those observed in FSRQs. In BL Lacs, on the other hand,
low values of $\dot m$ sustain intermediate or low disk
luminosities, then comparable to those directly extractable from a
Kerr hole after eq. (3). This yields a limiting power of about
$10^{46}$ erg s$^{-1}$; it is intrinsic to the present view that
the top BL Lac outputs should not considerably exceed such values,
as in fact observed (see Maraschi 2001).

c) SEDs: also the main features of the Blazar spectra can be
predicted in this picture as shown in \S 4. Comparing  BL Lacs
with FSRQs, in the former sources we expect lower particle
densities in the acceleration region, and less effectively
screened electric fields. These produce higher electron energies
up to some $10^3$ GeV, allowed on the other hand by the lower
photon densities; the ensuing synchrotron emission will peak at
higher frequencies. From b) and c) we derive the inverse
correlation $\nu_{peak} \propto L_d^{-1}$ or somewhat steeper, see
\S 4. But in BL Lacs specific and considerable scatter is to arise
when one tries to correlate  $\nu_{peak}$ with  the total
luminosity $L$, because this is made up by two comparable
components, $L_K$ and $L_d$.

d) Evolution: comparing FSRQs with BL Lacs, much weaker evolution
is expected for the latter, as argued in \S 5. This is because
their output is lower and is partly extracted from the Kerr hole
rotational energy, in conditions of low and long lasting accretion
rates; so these  sources are are long lived, with a slow
population evolution of the LE type over scales $\tau_L $ of
several Gyrs. Using only this scale to predict the source counts
from the relations given in \S 5,  the result is $N(>F) \propto
F^{-1.6}$ at fluxes $ F > 10^{-2}$ Jy. In addition, we expect a
genetic link to also occur; some FSRQs may make a transition to
the BL Lac mode after a ``last interaction" of their host with
companion galaxies. Such BL Lac births are equivalent to a
negative, though weak DE on a scale $\tau_D \approx  - 6$ Gyr;
this competes with the slow LE to yield counts flatter yet, namely
$N(> F) \propto F ^{-1.5}$ at intermediate fluxes, as shown fig.
1; equivalently, we expect values of $V_{e}/V_{a} \approx 0.5$,
see Giommi, Menna \& Padovani 1999. This scenario implies {\it
transitional} objects to occur from FSRQs to BL Lacs, with $\dot m
\mincir 10^{-1}$ and shining with intermediate luminosities for a
few $10^{-1}$ Gyr; in the optical band these sources will still
show thermal features (emission lines and bumps), while their
non-thermal SEDs will peak at higher $\nu_{peak}$ compared with
canonical FSRQs. On the other hand, similar relations computed
with the shorter time scales appropriate for the FSRQs yield much
steeper counts, as shown in fig. 2.

We note that {\it if} the radiosources FRI and II  constitute the
parent populations of the BL Lacs and the FSRQs, respectively,
then a genetic link between BL Lac births and FSRQ deaths implies
a similar link between the related radiosurces. Such a link
constitutes matter of current  debate. We recall from \S 2 that
power levels and optical spectra of the FRII and FRI radiosources
have been related to high and low values of $\dot m$,
respectively, by Baum, Zirbel {\&} O'Dea 1995. We stress that
overlapping properties and truly transitional objects are being
found (see Bicknell 1995, and PU01 with references therein); in
addition, similar evolutionary behaviors are observed for the
radiosources and for the Blazars related to similar values of
$\dot m$, that is to say strong evolution for the FRII but weak or
even negative evolution for the FRI radiosources (Jackson \& Wall
2001). So there is scope for discussing ``grand unification''
schemes of radiosources and Blazars, as pointed out by PU01.

We also note that our BMS agrees well with the parallel work by
B\"ottcher \& Dermer 2001; this is based on detailed modeling of
the Blazar emitting region, and concerns the full SEDs evolving
from FSRQs to BL Lacs as the accretion rates decrease and the
black hole environments are depleted in gas, dust and reprocessed
photons. In fact, the two papers turn out to be complementary, and
concur toward the notion of a sequence or even a genetic link
between the two subclasses.

Finally, if we carry the above BMS to its extreme, that is, to
residual $\dot m \mincir 10^{-4} $, we are led to expect objects
with very faint, if any, e.m. emission (further suppressed in the
ADAF regime, see Di Matteo et al. 2000), along with nearly
unscreened electric fields; these objects would be akin to the
cosmic ray accelerators discussed by Boldt {\&} Ghosh (1999).

Under the widespread if arguable assumption that in such
conditions a magnetic field $B \sim 10^2 - 10^3$ G can still be
held around the hole with the slow upward decrease recalled in \S
4, particles including protons would be accelerated over long
distances, at radii that must exceed the force-free region's
(acceleration at small radii has been shown to be untenable by
Levinson 2000).
 These particles -- in fact, cosmic rays with  ultra high
energies -- are predicted by the analogous of eq. (5) but in
conditions of unscreened fields, that is, large $d$ due to very
low $n$ (and to consistently large $\gamma$). Thus we recover the
long recognized energy limit given by
$${\cal E}_{max} \sim e \, B_d \, r_{ms}\,(r_{ms}/r)^p/p \sim 10^{20}\,
  B_3\, M_9^{5/4} \, r_{16}^{-1/4} \; eV ~. \eqno (11)$$

Tens of these accelerators could lie within 50 Mpc. As noted by
Boldt {\&} Ghosh (1999), they  would evade in the simplest way the
GZK cutoff at UHE; as to accounting for the observed particle
flux, they need to produce only $L \sim 10^{42}$ erg s$^{-1}$
mostly in energetic particles. Intergalactic magnetic fields of nG
strength would blur to near isotropy the geometrical memory of the
actual sources for most such UHECRs.

\smallskip
\noindent {\bf Acknowledgements:} We are indebted to P. Giommi, L.
Maraschi, P. Padovani and M. Salvati for several helpful
discussions. We thank our referee for useful and constructive
comments. Partial grants from ASI and MIUR are acknowledged.

\vfill\eject

\newpage

\medskip
%\vspace{0.2cm} %TO ALLOW SUFFICIENT SPACE BETWEEN THE TEXT AND THE FIGURES
\centerline{\bf Table 1 - The Blazar Main Sequence summarized}
\begin{table}[h]
\hspace{1.0cm} %if you want to center your table act on this argument
\begin{tabular}{|l|c|c|c|}
%\multicolumn{8}{|c|}{Month} \\
\hline
          &FSRQs  &BL Lacs & $\rightarrow$ CR accelerators \\
\hline key parameter       &$\dot m \sim 1$      &$\dot m \sim
10^{-2}$
   &$\dot m \mincir 10^{-4}$     \\

\hline
optical features    & em. lines, bump & $\sim $ no lines and bump  & none \\
\hline integrated power      &$L \sim 10^{47}$ erg s$^{-1}$
        &$L \mincir 10^{45}$ erg s$^{-1}$ &
        $L \mincir 10^{42}$ erg s$^{-1}$ \\
\hline Kerr hole vs. disk           &$L_K \ll L_d$     & $L_{K}
\mincir L_d$
                           & $L_{K} > L_d$ \\
\hline

top energies              & $h\nu \sim 10$  GeV   & $h\nu \sim 10$
TeV
            &  ${\cal E}_{max}\sim 10^{21}$ eV  \\
\hline

evolution             & strong & weak if any & negligible \\

\hline
%Kerr hole vs. disk           &$L_K \ll L_d$     & $L_{K} \mincir L_d$
%                           & $L_{K} > L_d$ \\
%\hline
%key parameter: $\dot m$      &$\dot m \sim 1$      &$\dot m \sim 10^{-2}$
%   &$\dot m \mincir 10^{-4}$     \\
%\hline
\end{tabular}
%%%%%%%%%%%%%%%%%%%%%%%%%%%%%%%%%%%%%%%%%%%%%%%%%%%%%%%%%%%%%%%%%%%%%%%%%
\end{table}
%%%%%%%%%%%%%%%%%%%%%%%%%%%%%%%%%%%%%%%%%%%%%%%%%%%%%%%%%%%%%%%%%%%%%%%%%%

\newpage
\section*{Figure Captions}
\vskip 2cm

Fig. 1. The BL Lac counts evaluated from eqs. (8) and (9) using
$\tau_D = - \, 7$ Gyr (dashed) and $\tau_D = -\, 5 $ Gyr (solid).
The dotted line represents the Euclidean counts. The data points
concern HBLs and are taken from Giommi et al. 1999, see their fig.
5.

Fig. 2. The FSRQ counts evaluated from eqs. (8) and (9) using
$\tau_D = + \, 6$ Gyr and $\tau_L = \, 3$ Gyr (solid line). The
open squares represent FSRQ counts from Wall \& Jackson 1997.

\newpage
\begin{figure}
\epsfysize=15cm 
\hspace{0.5cm}\epsfbox{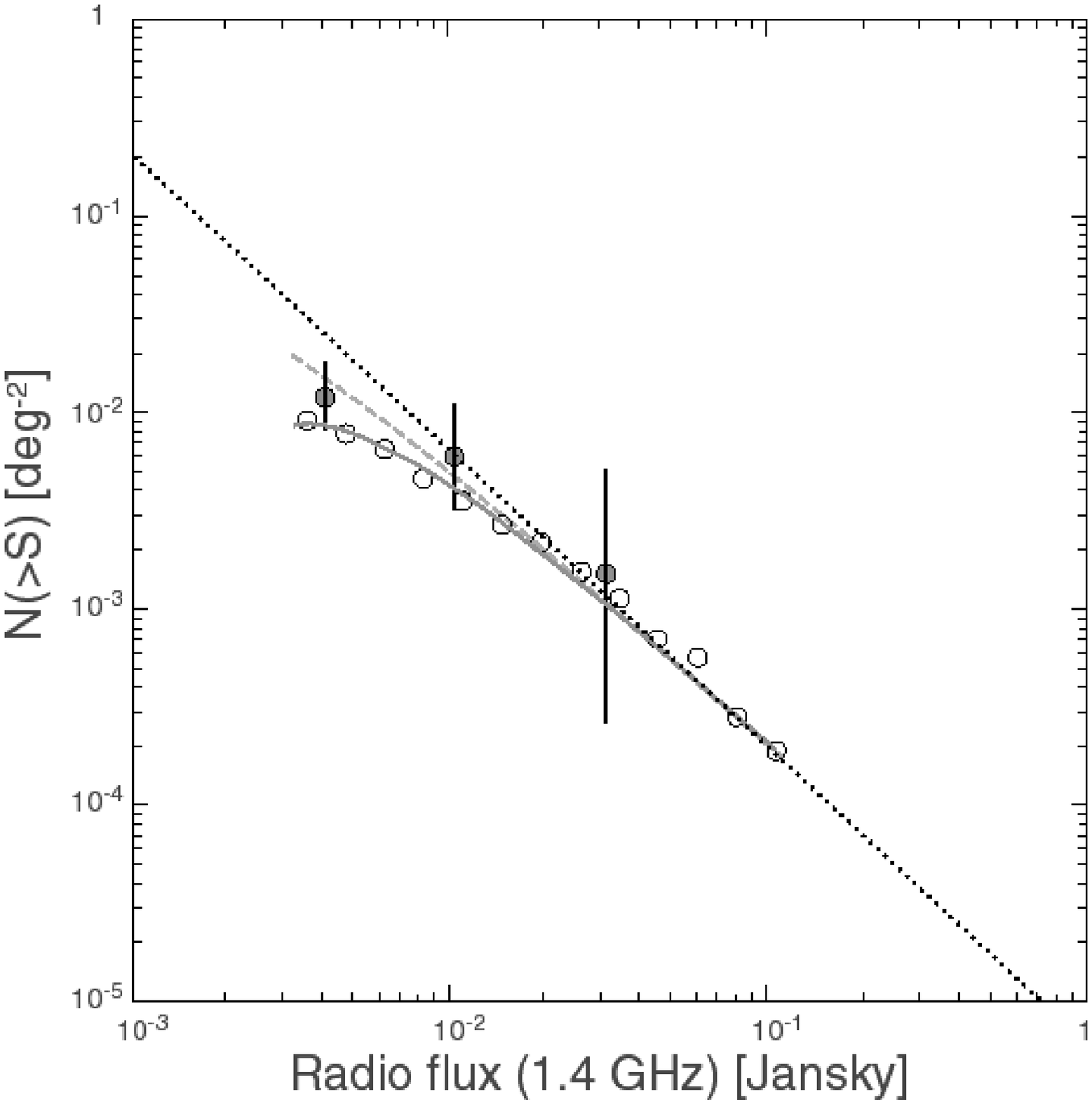} 
\end{figure}  

\newpage
\begin{figure}
\epsfysize=15cm 
\hspace{0.5cm}\epsfbox{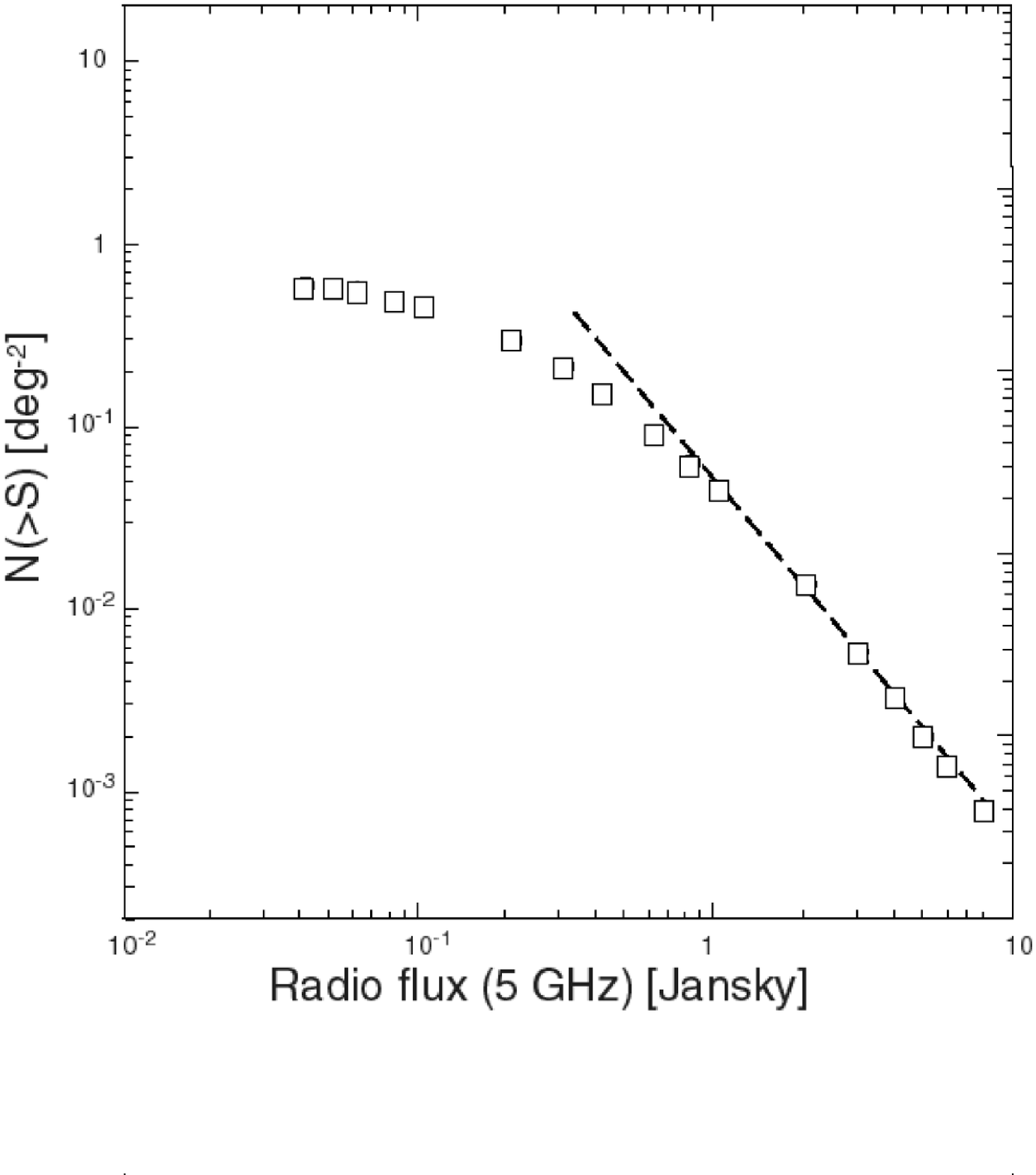} 
\end{figure}  

\end{document}